\documentclass{PoS}

\newcommand{\pslash}{p\hspace{-2mm}\slash}
\newcommand{\vecpslash}{\vec p\hspace{-2mm}\slash}
\newcommand{\A}{\mathcal{A}}
\newcommand{\B}{\mathcal{B}}

\newcommand{\intp}{\int dp_0\,}

\title{Propagators in lattice Coulomb gauge and confinement mechanisms}

\ShortTitle{Propagators in lattice Coulomb gauge and confinement mechanisms}

\author{\speaker{Giuseppe Burgio}\\
        E-mail: \email{giuseppe.burgio@uni-tuebingen.de}}
\author{Markus Quandt
}
\author{Mario Schr\"ock%
        \thanks{Current address: Karl-Franzens-Universit\"at, Graz, Austria.}
}
\author{Hugo Reinhardt\\
        Institut f\"ur Theoretische Physik, Eberhard-Karls-Universit\"at,
T\"ubingen, Germany}

\abstract{We discuss the gluon propagator in 3- and 
4-dimensional Yang-Mills theories in Coulomb gauge and compare it with the 
corresponding Landau
gauge propagator, showing that for both the relevant IR mass scale coincides.
We also report preliminary results on Coulomb gauge ghost form factor and quark propagators
and give a comment on the gluon propagator's strong coupling limit.}

\FullConference{The XXVIII International Symposium on Lattice Field Theory\\
		 June 14-19,2010\\
		 Villasimius, Sardinia Italy}

\begin{document}

\section*{Introduction}
\label{sec:introduction}
Yang-Mills theories wave functional studies in Coulomb gauge have enjoyed a renewed interest in the last years \cite{Szczepaniak:2001rg,Feuchter:2004mk,Schleifenbaum:2006bq,Epple:2006hv,Quandt:2010yq}, since they bypass an explicit construction of the physical Hilbert space \cite{Burgio:1999tg}. The Gribov-Zwanziger confinement scenario plays a key role in this context \cite{Gribov:1977wm,Zwanziger:1995cv} and its predictions for the static two pint functions need therefore to be explicitly verified. We report here on our lattice calculation program regarding static gluon and ghost, the relation between Coulomb and Landau gauge and the gluon strong coupling limit. We also show preliminary results for the quark propagator in Coulomb gauge, which should relate the Gribov-Zwanziger mechanism to chiral symmetry breaking.

\section{Gluon propagator}
\label{sec:gluon}

\begin{figure}[h]
\begin{center}
\includegraphics[width=0.6\textwidth,height=0.4\textwidth]{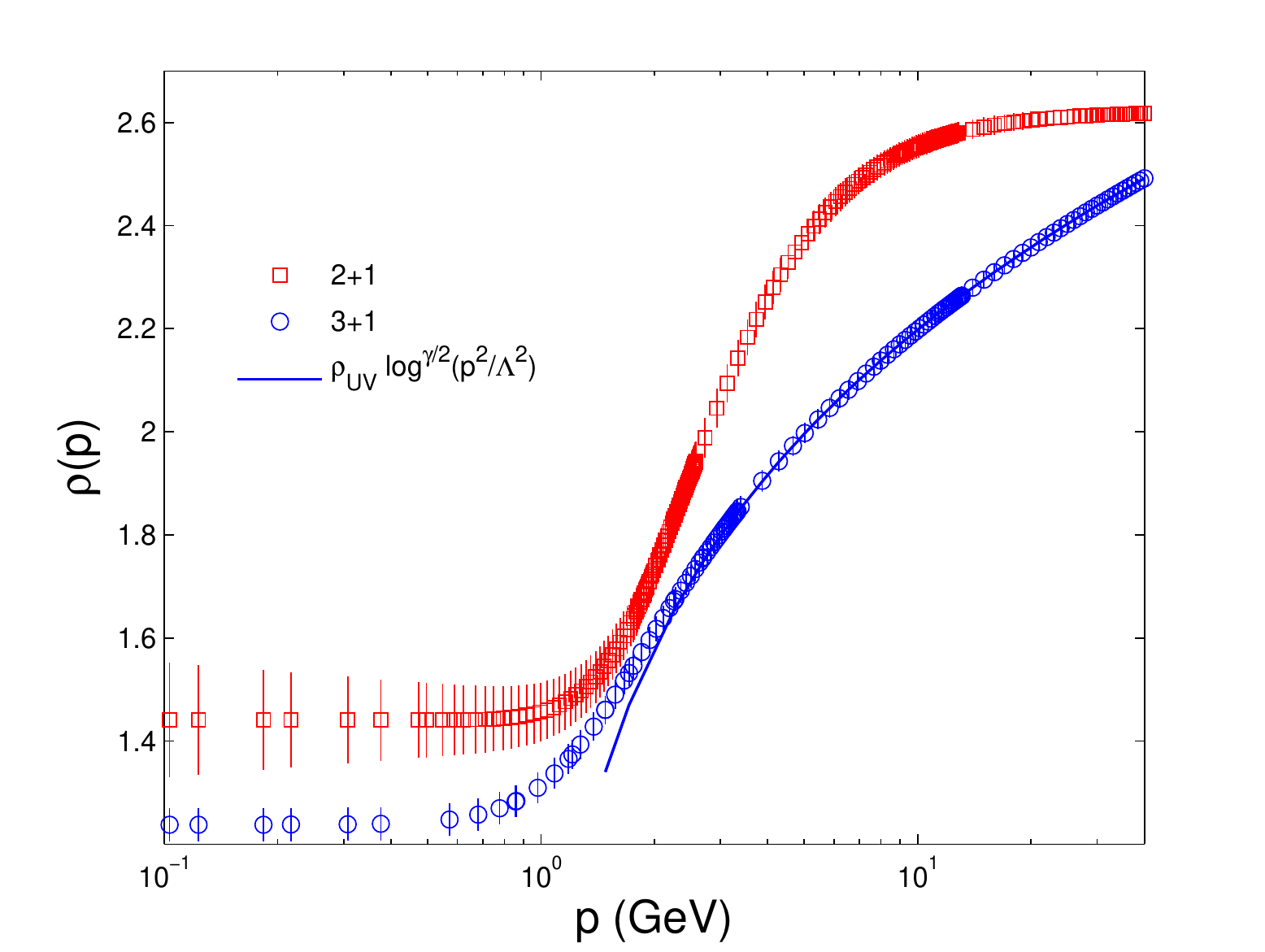}
\end{center}
\caption{Momentum rescaling function between Coulomb and Landau gauge propagators.}
\label{fig1}
\end{figure}

\begin{figure}
\includegraphics[width=0.45\textwidth,height=0.35\textwidth]{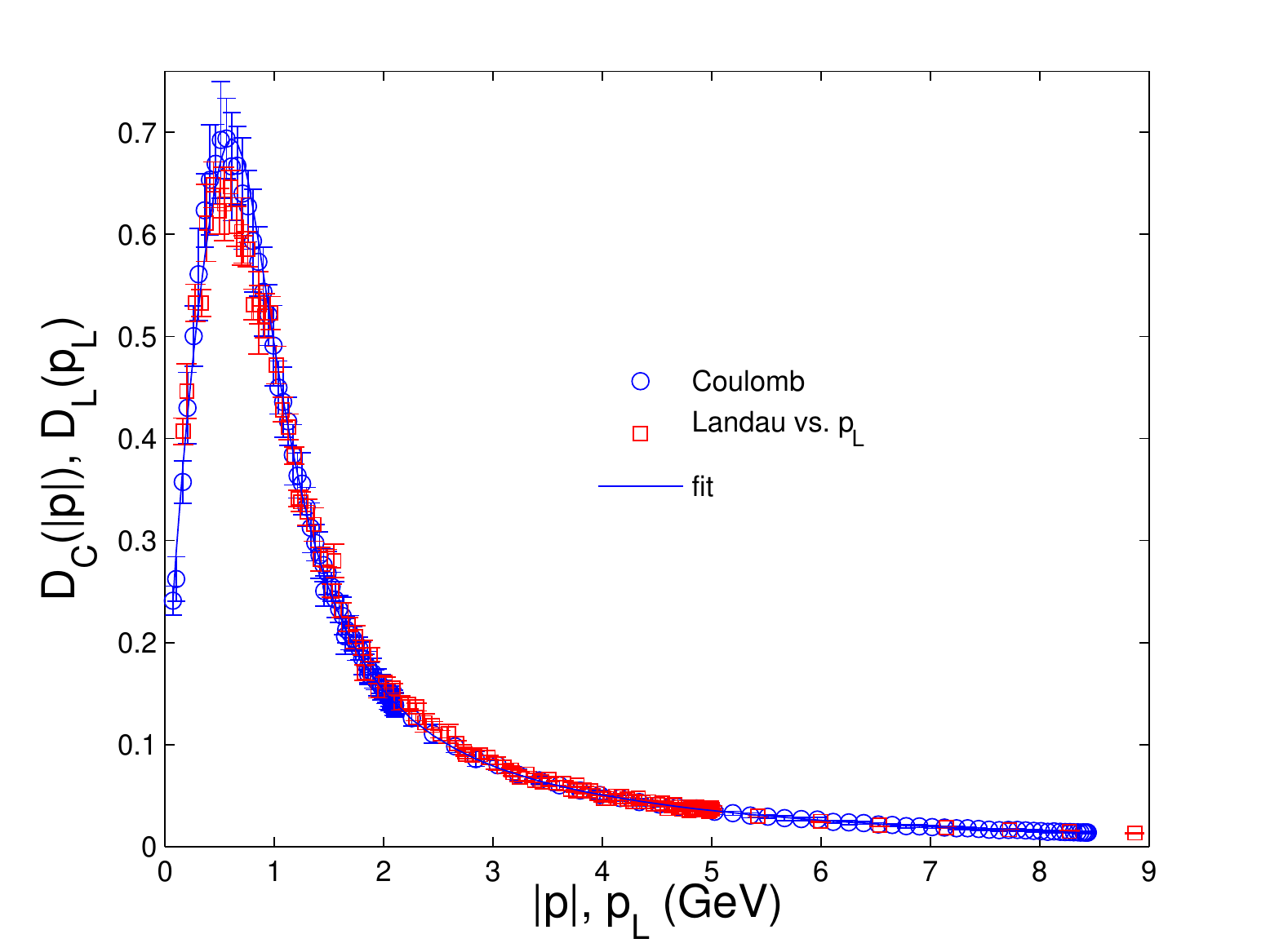}
\includegraphics[width=0.45\textwidth,height=0.35\textwidth]{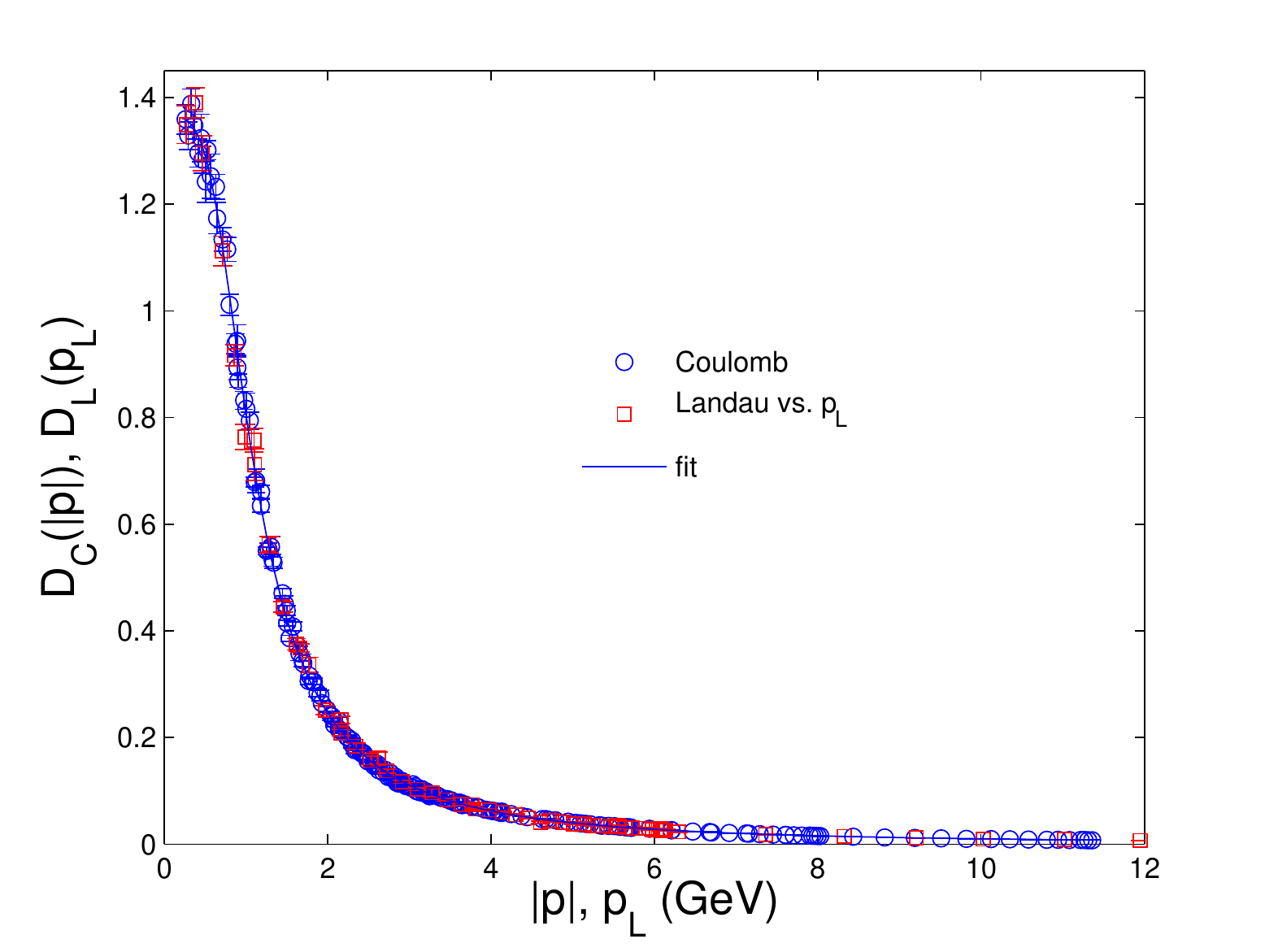}
\caption{Landau and Coulomb gluon propagator vs. equivalent momenta in 3 (left) end 4 dim. (right).}
\label{fig2}
\end{figure}

As shown in \cite{Burgio:2008jr,Burgio:2008yg}, after taking care of subtle renormalization issues related to the energy dependence, the $SU(2)$ static gluon propagator in Coulomb gauge $D(|\vec{p}|)$ agrees within numerical precision with Gribov's formula \cite{Gribov:1977wm}, with an IR mass $M=0.856(8) {\mbox{GeV}}$ and vanishing anomalous dimension in the UV. In  \cite{Burgio:2009xp} the relation between Coulomb and Landau gauge gluon propagators was futher investigated. It was shown that both in 3 and 4 dimensions the two propagators coincide, up to a simple rescaling $\rho(p)$ of the momentum variable, shown in Fig.~\ref{fig1}. 
The rescaling function in 3 dimensions is a simple interpolation between two constants, since both propagators have no anomalous dimension. In 4 dimensions only the Landau propagator has an anomalous dimension; the function $\rho$ behaves accordingly. When redrawn as a function of $p\,\rho(p)$, as shown in Fig.~\ref{fig2}, the Landau gluon propagator $D_L(p)$ coincides with the equivalent Coulomb one, $D_C(\vec{p})  = |\vec{p}|^{-1} D(|\vec{p}|)$; both show a massive IR behavior. Whether dimension 2 condensates \cite{Burgio:1997hc,Boucaud:2000ey} could be related to such IR mass is still an open question.

\section{Strong coupling limit}
\label{sec:strong}

\begin{figure}[h]
\begin{center}
\includegraphics[width=0.6\textwidth,height=0.4\textwidth]{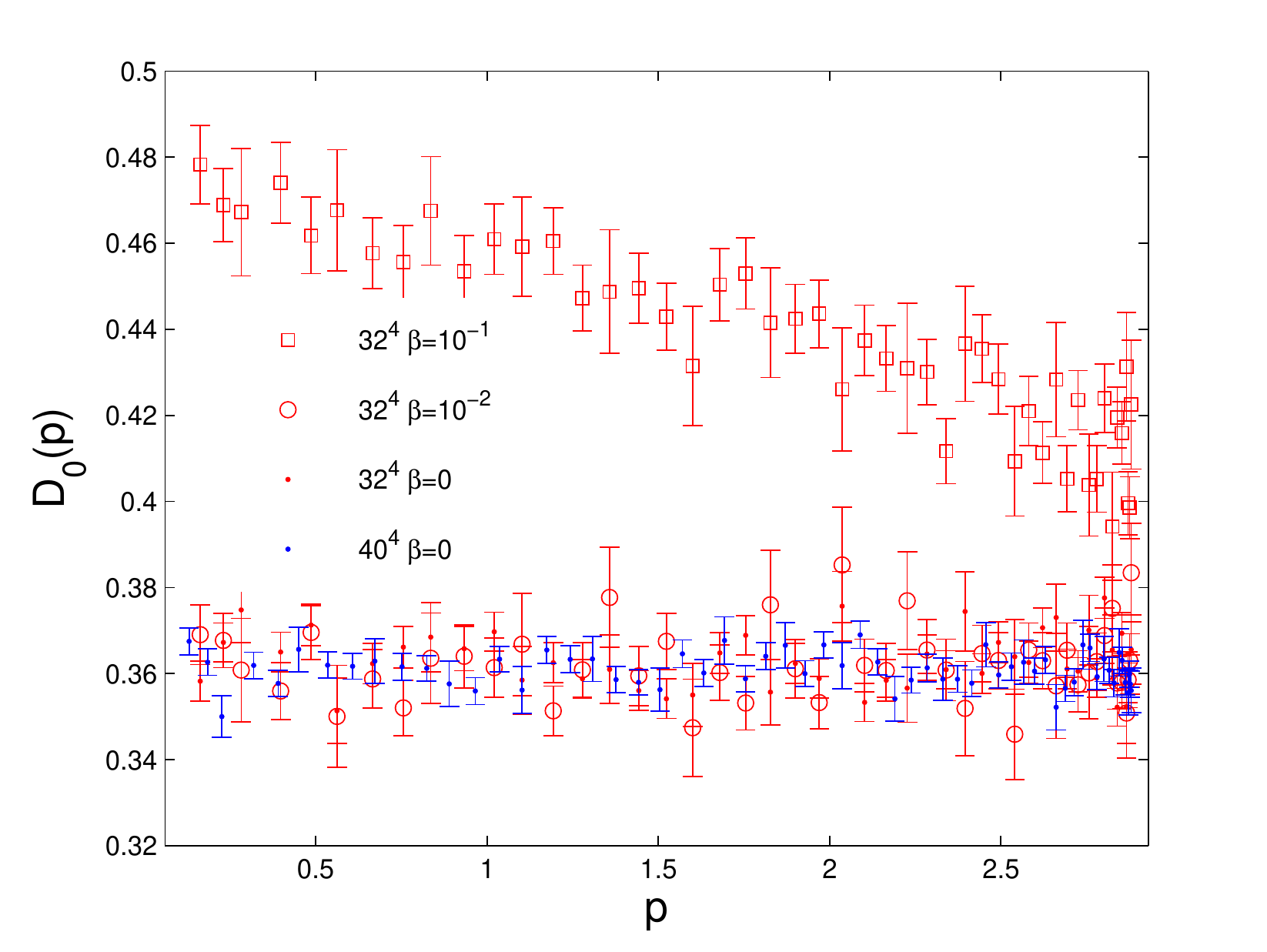}
\end{center}
\caption{Coulomb gauge temporal gluon propagator $D_0(|\vec{p}|)$ in the strong coupling limit.}
\label{fig3}
\end{figure}

The massive IR behavior in Landau gauge, found in all standard lattice calculations \cite{Bogolubsky:2005wf,Bogolubsky:2007bw}, is in disagreement with the Kugo-Ojima picture \cite{Kugo:1979gm}. Some authors \cite{Sternbeck:2008mv,Maas:2009ph} have supposed that discretization and/or Gribov copies effects might be responsible for the discrepancies; strong coupling calculations have been put forward as evidence. However going away from the continuum limit one goes through bulk transitions \cite{Burgio:2007np} and unphysical effects should be expected, so that predictions arising from $\beta=0$ results should be taken with extreme care. As an example, we show the temporal gluon propagator in Coulomb gauge $D_0(|\vec{p}|)$, which should be equivalent to the static potential \cite{Zwanziger:1995cv}. Even without going to anisotropic lattices \cite{Burgio:2003in}, which take care of scaling violations \cite{Burgio:2008jr,Burgio:2008yg,Nakagawa:2009is}, its diverging behavior, as expected for a confining theory, has been long established in the continuum limit. In Fig.~\ref{fig3} $D_0(|\vec{p}|)$ is given in lattice units; the physical behavior seems to be completely lost at strong coupling. The only way out would be to accept that the lattice scale diverges in the $\beta\to 0$ limit. All data would then collapse to a single point at $(0,\infty)$ when expressed in physical units. The physical information contained in such result is of course minimal.

\section{Ghost form factor}
\label{sec:ghost}

\begin{figure}[h]
\begin{center}
\includegraphics[width=0.6\textwidth,height=0.4\textwidth]{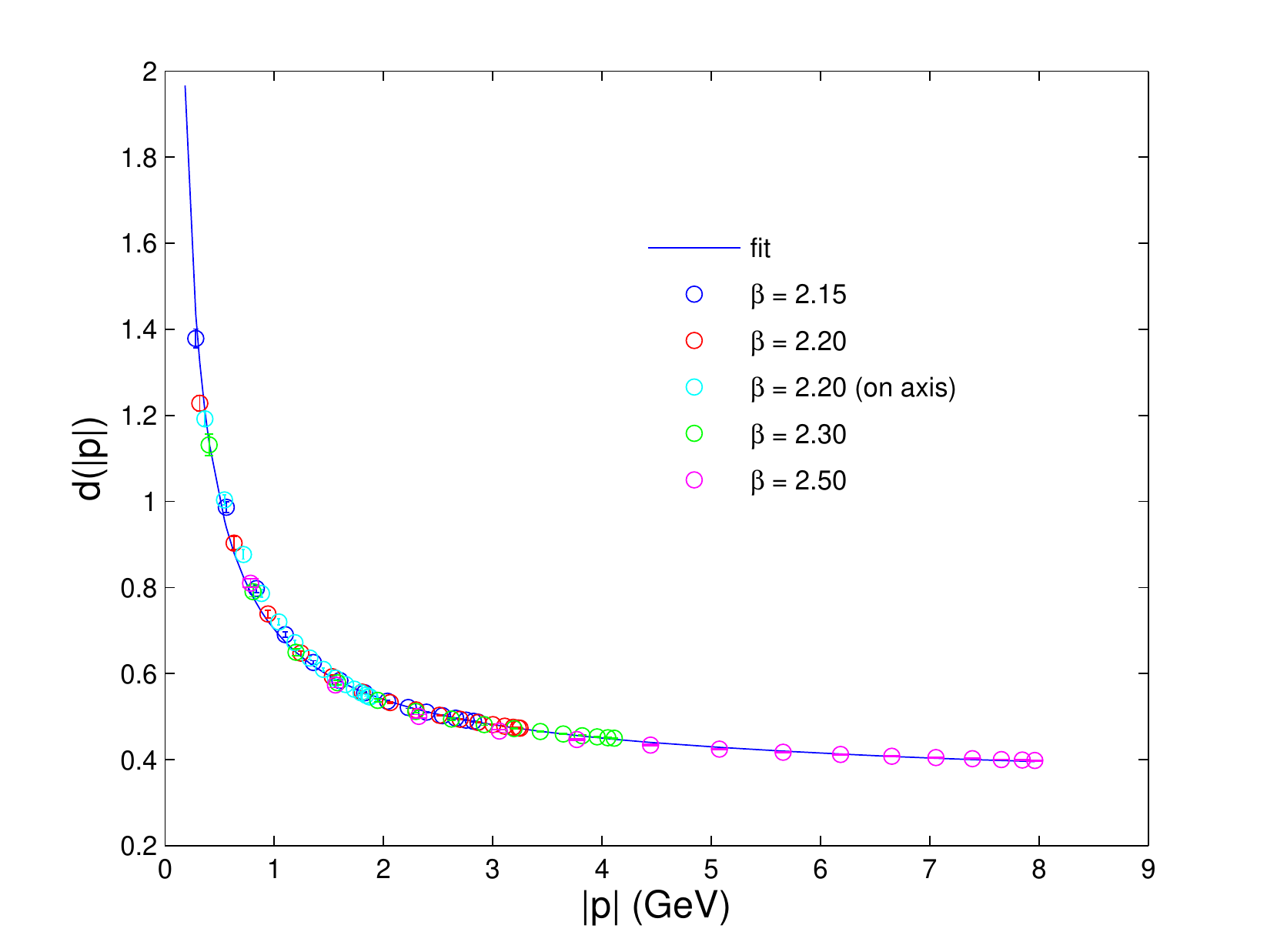}
\end{center}
\caption{Ghost form factor $d(|\vec{p}|)$ on a $32^4$ lattice.}
\label{fig4}
\end{figure}

Functional methods \cite{Schleifenbaum:2006bq,Epple:2006hv} provide sum rules which must be satisfied by the static gluon propagator and ghost form factor IR exponents $\kappa_{gl} + 2 \kappa_{gh} = 1$ and UV anomalous dimensions, $\gamma_{gl} + 2 \gamma_{gh} = 1$, and give precise predictions for their value, $\kappa_{gl} = -1$,  $\kappa_{gh}=1$, $\gamma_{gl} = 0$ and $\gamma_{gh} = 1/2$. For the static gluon propagator these have been already verified \cite{Burgio:2008jr,Burgio:2008yg,Burgio:2009xp}. Extending previous $SU(2)$ calculations \cite{Quandt:2007qd}, new preliminary results for the ghost form factor $d(|\vec{p}|)$ on a $32^4$ lattice, shown in Fig.~\ref{fig4}, seem to confirm such prediction. Indeed, we find that the data are very well fitted in the whole momentum range by:
\begin{equation}
d(|\vec{p}|) \propto \sqrt{\frac{m^2}{\vec{p}^2} + \log^{-1} (e + \frac{\vec{p}^2}{m^2})}\,,
\end{equation}
with $m=0.32(2) {\mbox GeV}$ and $\chi^2$/d.o.f. $= 0.8$. Calculations on larger lattices are however necessary and under way to confirm this result \cite{Markus2010}.

\section{Quark propagator}
\label{sec:quark}

\begin{figure}[h]
\begin{center}
\includegraphics[width=0.6\textwidth,height=0.4\textwidth]{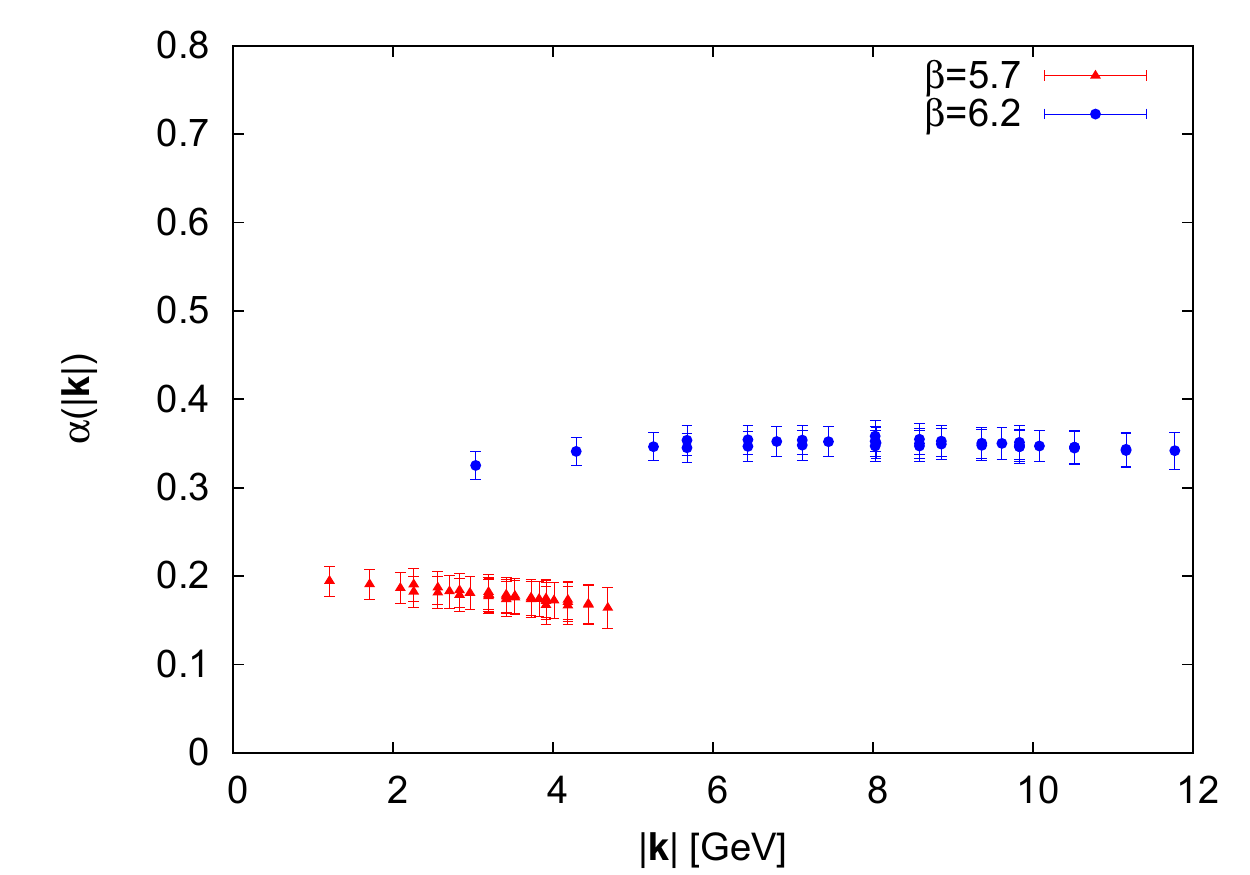}$\quad$
\end{center}
\caption{Temporal component $\alpha(\vec{k},k_0)$ of the quark propagator in Coulomb gauge.}
\label{fig5}
\end{figure}

\begin{figure}
\includegraphics[width=0.45\textwidth,height=0.35\textwidth]{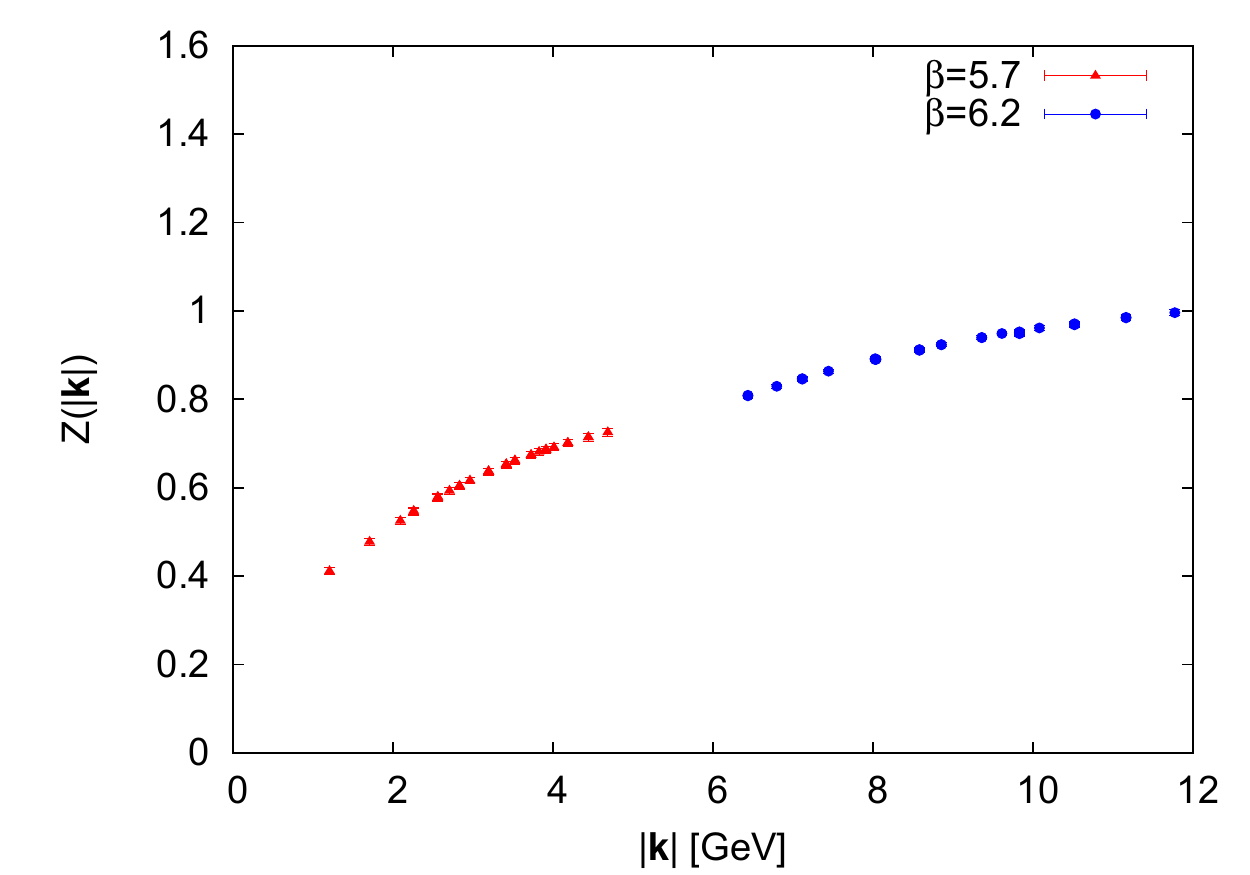}
\includegraphics[width=0.45\textwidth,height=0.35\textwidth]{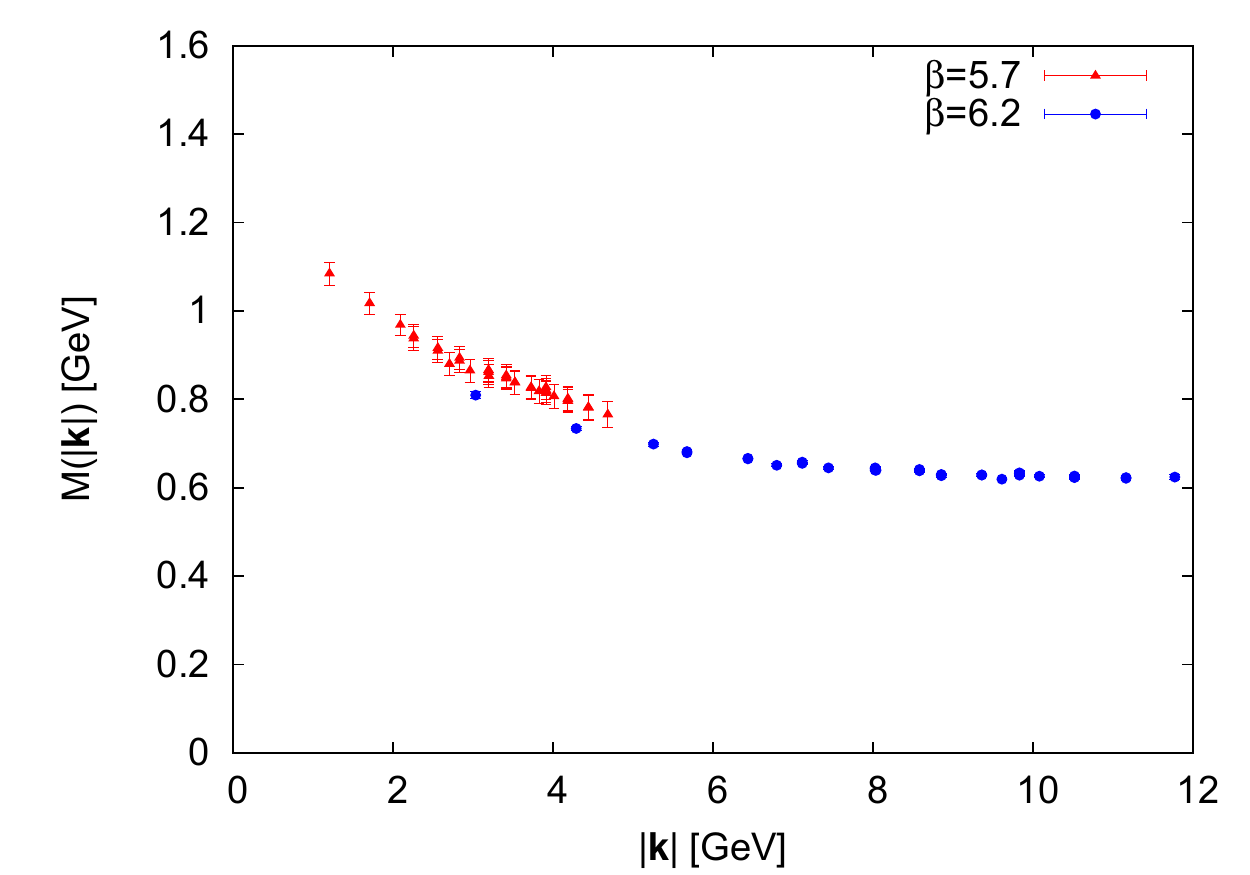}
\caption{Static quark self energy $Z(\vec{k})$ and running mass $M(\vec{k})$.}
\label{fig6}
\end{figure}

Although not directly related to the Gribov-Zwanziger confinement scenario, the quark propagator in Coulomb gauge provides important physical information on chiral symmetry breaking and dynamical mass generation. We report here preliminary results obtained in $SU(3)$ for quenched $12^3 \times 24$ lattices, using asqtad staggered fermions with a bare mass of $m_b \simeq$ 212 MeV. Results using unquenched configurations and masses closer to the chiral limit will be published soon \cite{Mario2010}. We have followed \cite{Skullerud:2000un,Bowman:2002bm,Bowman:2005vx}, adapting their method and notation from Landau to Coulomb gauge, and explicitly inverted the Dirac operator on each configuartion. Our first non-trivial results is that the coefficients of the mixed term $\pslash_0\,\vecpslash$, which are in principle allowed in Coulomb gauge beyond tree level, vanish within numerical precision. We find then that the most general form for the propagator and its inverse reads:
\begin{eqnarray}
S(\vec{p},p_0) &=& -i\, \pslash_0\,  \A_t(\vec{p},p_0) - i\, \vecpslash\,  \A_s(\vec{p},p_0)   + \B_m(\vec{p},p_0)\nonumber\\
S^{-1}(\vec{p},p_0) &=& \,\;\;\;i\, \pslash_0 \,\, A_t(\vec{p},p_0) + i\, \vecpslash\, \,\, A_s(\vec{p},p_0) \,+\, B_m(\vec{p},p_0)\,,\label{prop}
\end{eqnarray}
where $A_t$, $A_s$ and $B_m$ can be easily re-expressed in terms of $\A_t$, $\A_s$ and $\B_m$, which are explicitly obtained from the simulations.  If renormalizable, one should expect:
\begin{eqnarray}
S(\vec{p},p_0) &=& Z(\vec{p},p_0)\;\; \frac{1}{i\vecpslash  + i\pslash_0 \alpha(\vec{p},p_0)+M(\vec{p},p_0)}\nonumber\\
S^{-1}(\vec{p},p_0) &=& \frac{1}{Z(\vec{p},p_0)} \left[ i\vecpslash 
      + i\pslash_0 \alpha(\vec{p},p_0)+M(\vec{p},p_0)\right]
\end{eqnarray}
and $Z(\vec{p},p_0)$, $\alpha(\vec{p},p_0)$ and $M(\vec{p},p_0)$ can be again expressed in terms
of $\A_t$, $\A_s$ and $\B_m$; $Z(\vec{p},p_0)$ should be multiplicatively renormalizable while the functions $\alpha(\vec{p},p_0)$ and $M(\vec{p},p_0)$ should be cut-off independent. Fig.~\ref{fig5} shows the former for two different values of the coupling; it is clear that the full quark propagator $S(\vec{p},p_0)$ is not renormalizable. This seems to be a general pattern: just like for the full Coulomb gauge gluon propagator \cite{Burgio:2008jr,Burgio:2008yg,Burgio:2009xp}, the $p_0$ dependence needs to be treated separately.
Turning to the static quark propagator $S^{-1}(\vec{p}) = \int d\,p_0 S^{-1}(\vec{p},p_0)$, being $\alpha$ an even function of $p_0$, the temporal component cancels for parity and we obtain:
\begin{eqnarray}
S(\vec{p}) &=& {Z(\vec{p})}\; \frac{1}{i\vecpslash +M(\vec{p})}\nonumber\\
S^{-1}(\vec{p}) &=& \frac{1}{Z(\vec{p})} \left[ i\vecpslash +M(\vec{p})\right]\,,
\end{eqnarray}
where:
\begin{eqnarray}
M(\vec p) &=& \frac{\intp B_m(\vec{p},p_0)}{\intp A_s(\vec{p},p_0)} =\frac{\intp \B_m(\vec{p},p_0)}{\intp \A_s(\vec{p},p_0)}\nonumber\\
Z(\vec p) &=& \frac{1}{\intp A_s(\vec{p},p_0)} \;=\left[\vec p^2+M^2(\vec p) \right] \intp \A_s(\vec{p},p_0)\,.
\label{ZM}
\end{eqnarray}
$Z(\vec{p})$ is proportional to the quark dispersion relation and should be multiplicative renormalizable, while the running mass $M(\vec{p})$ should be renormalization group invariant. The two functions are shown in Fig.~\ref{fig6}, where $Z(\vec{p})$  has been rescaled for different $\beta$ while the values for $M(\vec{p})$ have been directly taken from Eq.~\ref{ZM} without modifications; the light discrepancy is simply due to the different renormalized current mass for different cutoffs given a fixed bare mass $m_b$. A cone cut has been applied to both data sets.

\section*{Conclusions and Outlook}
\label{sec:conclusions}

We have reported on the progress in our program to determine two point functions in Coulomb gauge. The static gluon propagator and ghost form factor confirm down to the available IR momenta the Gribov-Zwanziger confinement scenario and the predictions from the hamiltonian functional approach. We also show that, as for the gluon, only the static component of the quark propagator is renormalizable. Extension to higher volumes and unquenched configurations are currently under investigation \cite{Markus2010,Mario2010}. We also comment on strong coupling limit results for the gluon propagator. We show that, as expected, physical interpretations in this limit must be taken with extreme care.

\acknowledgments

We thank D. Campagnari and P. Watson for helpful discussions. This work was partly supported by DFG under the contract Re856/6-3.


\begin{thebibliography}{99}

\bibitem{Szczepaniak:2001rg}
{{A.~P.} {Szczepaniak}}
  {and} {{E.~S.}
  {Swanson}}, \emph{Phys. Rev.}
  \textbf{{D65}}, {025012}
  ({2002}), [{\tt hep-ph/0107078}].

\bibitem{Feuchter:2004mk}
{{C.}~{Feuchter}} {and}
  {{H.}~{Reinhardt}},
  \emph{Phys. Rev.} \textbf{{D70}},
  {105021} ({2004}), [{\tt hep-th/0408236}].

\bibitem{Schleifenbaum:2006bq}
{{W.}~{Schleifenbaum}},
  {{M.}~{Leder}} {and}
  {{H.}~{Reinhardt}},
  \emph{Phys. Rev.} \textbf{{D73}},
  {125019} ({2006}), [{\tt hep-th/0605115}].

\bibitem{Epple:2006hv}
{{D.}~{Epple}},
  {{H.}~{Reinhardt}} {and}
  {{W.}~{Schleifenbaum}},
  \emph{Phys. Rev.} \textbf{{D75}},
  {045011} ({2007}), [{\tt hep-th/0612241}].

\bibitem{Quandt:2010yq}
{{M.}~{Quandt}}, {{H.}~{Reinhardt}} {and} {{G.}~{Burgio}}, 
  \emph{Phys. Rev.} \textbf{{D81}},
  {065016} ({2010}), [{\tt arXiv:1001.3699}].

\bibitem{Burgio:1999tg}
{{G.}~{Burgio}}, {{R.}~{De Pietri}}, {{H.~A.}~{Morales-Tecotl}}, 
{{L.~F.}~{Urrutia}} {and} {{J.~D.}~{Vergara}}, 
  \emph{Nucl. Phys.} \textbf{{B566}},
  {547-561} ({2000}), [{\tt hep-lat/9906036}].

\bibitem{Gribov:1977wm}
V.~N.Gribov, \emph{Nucl. Phys} \textbf{{B139}}, 1 (1978).

\bibitem{Zwanziger:1995cv}
{{D.}~{Zwanziger}},
  \emph{Nucl. Phys.} \textbf{{B485}},
  {185} ({1997}), [{\tt hep-th/9603203}].

\bibitem{Burgio:2008jr}
{{G.}~{Burgio}}, {{M.}~{Quandt}} {and} {{H.}~{Reinhardt}}, 
  \emph{Phys. Rev. Lett.} \textbf{{102}},
  {032002} ({2009}), [{\tt arXiv:0807.3291}].

\bibitem{Burgio:2008yg}
{{G.}~{Burgio}}, {{M.}~{Quandt}} {and} {{H.}~{Reinhardt}}, 
  \emph{{PoS}} \textbf{{{CONFINEMENT8}}},
  {051} ({2008}), [{\tt arXiv:0812.3786}].

\bibitem{Burgio:2009xp}
{{G.}~{Burgio}}, {{M.}~{Quandt}} {and} {{H.}~{Reinhardt}}, 
  \emph{Phys. Rev.} \textbf{{D81}},
  {074502} ({2010}), [{\tt arXiv:0911.5101}].

\bibitem{Burgio:1997hc}
{{G.}~{Burgio}},
  {{F.}~{Di~Renzo}},
  {{G.}~{Marchesini}}
  {and} {{E.}~{Onofri}},
  \emph{Phys. Lett.} \textbf{{B422}},
  {219} ({1998}), {hep-ph/9706209}.

\bibitem{Boucaud:2000ey}
{{P.}~{Boucaud}} {et~al.},
  \emph{JHEP} \textbf{{04}}, {006}
  ({2000}), {hep-ph/0003020}.

\bibitem{Bogolubsky:2005wf}
{{I.~L.} {Bogolubsky}}
  {et~al.}, \emph{Phys. Rev.}
  \textbf{{D74}}, {034503}
  ({2006}), [{\tt hep-lat/0511056}].

\bibitem{Bogolubsky:2007bw}
{{I.~L.} {Bogolubsky}}
  {et~al.}, \emph{Phys. Rev.}
  \textbf{{D77}}, {014504}
  ({2008}), [{\tt arXiv:0707.3611}].

\bibitem{Kugo:1979gm}
{T.}~{Kugo} {and}
  {{I.}~{Ojima}},
  \emph{Prog. Theor. Phys. Suppl.} \textbf{{66}},
  {1} ({1979}).

\bibitem{Sternbeck:2008mv}
{{A.}~{Sternbeck}} {and} {{L.}~{von Smekal}},
\emph{Eur. Phys. J.} \textbf{{C68}}, {487} ({2010})
 [{\tt arXiv:0811.4300}].

\bibitem{Maas:2009ph}
{{A.}~{Maas}}, {{J.~M.}~{Pawlowski}}, {{D.}~{Spielmann}}, {{A.}~{Sternbeck}} {and} {{L.}~{von~Smekal}},
\emph{Eur. Phys. J.} \textbf{{C68}}, {183} ({2010})
 [{\tt arXiv:0912.4203}].

\bibitem{Burgio:2007np}
{{G.}~{Burgio}}, 
  \emph{{PoS}} \textbf{{LAT2007}},
  {292} ({2007}), [{\tt arXiv:0710.0476}].

\bibitem{Burgio:2003in}
{{G.}~{Burgio}} {et}
{{al.}} ({TrinLat}),
  \emph{Phys. Rev.} \textbf{{D67}},
  {114502} ({2003}), [{\tt hep-lat/0303005}].

\bibitem{Nakagawa:2009is}
{{Y.}~{Nakagawa}},
  {{A.}~{Nakamura}},
  {{T.}~{Saito}} {and}
  {{H.}~{Toki}}
 \emph{{PoS}} \textbf{{LAT2009}},
  ({2009}), [{\tt arXiv:0911.2550}].

\bibitem{Quandt:2007qd}
{{M.}~{Quandt}},
  {{G.}~{Burgio}},
  {{S.}~{Chimchinda}}
  {and}
  {{H.}~{Reinhardt}},
  \emph{PoS} \textbf{{LAT2007}},
  {325} ({2007}), [{\tt arXiv:0710.0549}].

\bibitem{Markus2010}
{{M.}~{Quandt}},
  {{G.}~{Burgio}}
  {and}
  {{H.}~{Reinhardt}},
  \emph{In preparation}, ({2010}).

\bibitem{Mario2010}
 {{M.}~{Schr\"ock}},
 {{G.}~{Burgio}},
{{M.}~{Quandt}},
  {and}
  {{H.}~{Reinhardt}},
  \emph{In preparation}, ({2010}).
  
\bibitem{Skullerud:2000un}
 {{J.~I.}~{Skullerud}} {and} {{A.~G.}~{Williams}}, 
\emph{Phys. Rev.} \textbf{D63},
{054508} ({2001}), [{\tt hep-lat/0007028}].

\bibitem{Bowman:2002bm}
{{P.~O.}~{Bowman}}, {{U.~M.}~{Heller}} {and} {{A.~G.}~{Williams}},  
\emph{Phys. Rev.} \textbf{D66}
{014505} ({2002}), [{\tt hep-lat/0203001}].

\bibitem{Bowman:2005vx}
{{P.~O.}~{Bowman}} {et~al.}
\emph{Phys. Rev.} \textbf{D71}
{054507} ({2005}), [{\tt hep-lat/0501019}].
  
\end{thebibliography}
\end{document}